\documentstyle[12pt]{article}
\begin{document}

\def\half{{1 \over 2}}
\def\eee{{\rm e}}
\newcommand{\eeas}{\end{eqnarray*}}
\def\laa{\langle\kern-.3em \langle}
\def\raa{\rangle\kern-.3em \rangle}
\def\la{\lambda}
\def\cn{{\cal N}}\def\cz{{\cal Z}}
\def\ymt{Yang-Mills theory}
\def\sftH{string field theory Hamiltonian}
\def\aaa#1{a^{\dagger}_{#1}}
\def\aap#1#2{a_{#1}^{\dagger#2}}
\def\ee#1{\hbox{e}^{#1}}
\def\dd{\hbox{d}}
\def\tr{\hbox{tr}}
\def\Tr{\hbox{Tr}}
\def\part{\partial}
\def\refe#1{eq.~\ref{#1}}
\begin{titlepage}

\begin{flushright}
PUPT-1892 \\
hep-th/9909152 \\
\end{flushright}

\vskip 1.2 true cm

\begin{center}
{\Large \bf Dynamical truncation of the string spectrum at finite $N$ }
\end{center}

\vskip 0.6 cm

\begin{center}

Gilad Lifschytz and Vipul Periwal

\vspace{3mm}

{\small \sl Department of Physics} \\
{\small \sl Joseph Henry Laboratories} \\
{\small \sl Princeton University} \\
{\small \sl Princeton, NJ 08544, U.S.A.} \\
\smallskip
{\small \tt gilad@viper.princeton.edu, vipul@mail.princeton.edu}

\end{center}

\vskip 0.8 cm

\begin{abstract}
\noindent
We exhibit a nonperturbative background independent 
dynamical truncation of the string spectrum and a quantization of 
the string coupling constant directly from the Hamiltonian governing 
the dynamics of  strings constructed from Yang-Mills theories.  
\end{abstract}

\end{titlepage}

\section{Introduction}

The relation between Yang-Mills theories and string theories has been
a leitmotif in particle physics ever since Mandelstam\cite{sm}\ 
first noted the 
significance of Wilson loops as observables in Yang-Mills 
theories\cite{others,loop,jevics,pol}.  
Schwinger-Dyson equations satisfied by Wilson loop observables in the 
gauge theory were found to have natural interpretations in terms of 
geometric operations on strings, such as string splitting and 
joining\cite{loop}.  With $N$ the rank of the gauge group U$(N)$ in the \ymt, 
the topological classification of 
Feynman diagrams in the 't Hooft large $N$ limit\cite{thooft}\ in \ymt\ gave a
further reason for expecting that strings should have some 
significance in gauge theories. In this limit the  power of $N^{-2}$
accompanying a given Feynman diagram 
depends only on the topology of the surface defined by the double  
line representation of that  diagram.  The Schwinger-Dyson 
loop equations in the large $N$ limit also simplify---strings can no longer
join, and the equations are satisfied by the expectation values which 
factorize in this limit.  

The AdS$\leftrightarrow$CFT conjecture\cite{juan}\  
gives a geometric picture of the 
large $N$ limit of maximally supersymmetric \ymt\ 
in four dimensions. In a sense made precise in \cite{gkp,wmalda}\ it 
is 
equal to the classical type IIB supergravity theory compactified on 
${  AdS}_{5}\times S^{5}.$  For finite $N$ the correspondence is 
rather unclear.  A natural extension of the large $N$ conjecture to 
finite values of $N$ 
would be that the supergravity theory should be replaced by string 
theory on a background with the appropriate isometry group at finite 
$N.$  Given that the \ymt\ is perfectly 
well-defined at any value of $N$ and string theory on ${  AdS}_{5}\times S^{5}$
seems quite difficult to construct, it has been suggested that the 
\ymt\ is as good a definition of string theory on ${  AdS}_{5}\times S^{5}$
as one is likely to obtain.   

What of the Wilson loops that initiated the quest for a string 
description of \ymt\ in the first instance?  There is a 
natural string   theory canonically defined by the \ymt\ 
as follows\cite{v1}:
Define the loop operator
\begin{equation}
 H \equiv \int \ee{+S}{\delta\over{\delta A_{\mu}}} 
\ee{-S}{\delta\over{\delta A_{\mu}}}
\end{equation}
which acts on products of Wilson loops. The loop equations (which are 
the Schwinger-Dyson equations for Wilson loops) of the \ymt\ 
are then
\begin{equation}
\langle H\prod_{i} W(C_{i})\rangle = 0.
\end{equation}  
Here $\langle \ldots\rangle $ denotes correlation functions in the 
\ymt\ defined by a functional integral weighted by the action of the \ymt,
$S$.
This operator $H$ turns out to be the 
Fokker-Planck Hamiltonian\cite{marc}\ in Parisi-Wu stochastic 
quantization\cite{pw}\
of the \ymt.
This means introducing an additional Euclidean dimension
with a co\"ordinate $\tau $ (called the stochastic time).
In this enlarged co\"ordinate space,
an auxiliary classical stochastic theory is defined.  
The Hamiltonian  for evolution in 
the $\tau$ direction of this auxiliary  theory is 
just the operator $H.$ 
Correlation functions of Wilson loops in the \ymt\ are given by
boundary correlations  of the auxiliary theory: 
\begin{equation}
\langle \prod_{i} W(C_{i})\rangle = \lim_{\tau\uparrow\infty}
\laa 0 |\exp(-\tau H) \prod_{i} W(C_{i}) |0\raa,
\end{equation}
where $|0\raa$ denotes the ground state of the Hamiltonian $H.$
One now identifies the 
Fokker-Planck Hamiltonian\cite{jevic}\ with a  Hamiltonian of a noncritical 
string field theory in a very special gauge\cite{ik}---temporal gauge---where 
the  time ($\tau$) is identified with the Liouville degree of
freedom.  We say more about the string interpretation in our 
concluding comments.

We see therefore that  the Wilson loops and their Schwinger-Dyson 
equations of motion responsible for motivating a string interpretation 
of \ymt\ are also at the root of a direct identification of a string 
field theory from the gauge theory.

\section{Truncation of the spectrum}

Kaluza-Klein(KK) modes on   $S^5$ are associated with 
chiral primary operators in the supersymmetric 
\ymt, which are finite in number
at finite $N,$ due to simple matrix identities.
One particularly  interesting  aspect of the
AdS$\leftrightarrow$CFT correspondence is that it therefore implies that 
{\it a priori} independent excitations of the dual 
string theory  cannot be independent. However, the 
perturbative fundamental string does not  know about the 
integrality of $N.$

The aim of this paper is to exhibit the corresponding dynamical 
truncation
phenomenon directly in the string theories  
constructed as above from Yang-Mills 
theories.
 We shall show that the string Hamiltonian exhibits a
quantization of the string coupling (in our case it is just $1/N^2$)
 and a coupling constant dependent
truncation of the spectrum  of a single string, phenomena that 
are invisible in string perturbation theory to any order.

The simplest model of the noncritical string 
theory$\leftrightarrow$\ymt\ correspondence  is the one-plaquette 
model 
\begin{equation}
\cz = \int \dd U \ \exp\left((N/\la)\tr \left[ U  + 
 U^\dagger \right]\right).
\end{equation}
In our notation $\tr$ is the usual trace and $\Tr \equiv \tr/N.$  
Let $\{T_{\alpha}\}$ be a normalized anti-Hermitian orthonormal basis for 
the Lie algebra of  U$(N)$ and $h\equiv h^{\alpha}T_{\alpha}.$ 
As usual from the translation invariance of the U$(N)$ Haar
measure, we can derive 
Schwinger-Dyson equations for Wilson loops 
\begin{equation}
\sum_{\alpha} {\part\over{\part h^{\alpha}}} 
\int \dd U \ \ee{(N/\la) {\rm tr} \left[ U\eee^{{h}}  + 
 \eee^{{{-h}}}U^\dagger \right] }\sum_{j=1}^{m}n_{j}
\Tr\left(T_{\alpha}(U\ee{h})^{n_{j}}\right)
\prod_{i\not=j}\Tr\left((U\ee{h})^{n_{i}}\right)\biggm|_{h=0}
= 0 .
\label{ident}
\end{equation}
Interpreting this equation as the condition for equilibrium in
stochastic quantization 
\begin{equation}
 \lim_{\tau\uparrow\infty} {\dd\over\dd\tau}
\laa 0 | \exp(-\tau \hat H)  \prod_{i}\Tr U^{n_{i}} |0\raa =
\lim_{\tau\uparrow\infty}
\laa 0 | \exp(-\tau \hat H) [\hat H,\prod_{i}\Tr U^{n_{i}}]|0\raa = 0,
\end{equation}
one finds the   Hamiltonian\cite{v2}  
\begin{eqnarray}
\hat H &=&
 \sum_{n>0}n\bigg[ {1\over\lambda}
 \big(a^{\dagger}_{n+1}-(1-\delta_{n,1})a^{\dagger}_{n-1} - 
 \delta_{n,1}\big) 
 +a^{\dagger}_{n}+\sum_{p=1}^{n-1}
 a^{\dagger}_{p}a^{\dagger}_{n-p}\bigg] a_{n}\nonumber\\ 
&+& (\hbox{$n<0$ terms})
 +{1\over  N^{2}} 
\sum_{i,j}  ija^{\dagger}_{ {i}+ {j}}a_{i}a_{j}.
\label{ham}
\end{eqnarray}
Here $a_{i}^{\dagger}$ creates $\Tr U^{i}$ in the correlation 
function, $[a_{j},a_{i}^{\dagger}]=\delta_{i,j}, a_{0}^{\dagger}\equiv 1.$  
Thus we see explicitly the  weighting of joining interactions with a factor 
of $1/N^{2},$ while splitting interactions  are $O(1),$ keeping fixed 
the expectation value $\langle W(C)\rangle = O(1)$ in the large $N$ 
limit.  While $\hat H$ is not Hermitian, it is related to a Hermitian
Hamiltonian by a similarity transformation as usual\cite{pw}. 

The  raison d'etre of $\hat H$ is to reproduce the Schwinger-Dyson 
equations. In deriving $\hat H$ there is an implicit assumption that 
the $1/N^{2}$ is treated perturbatively, since no relationship
between traces of matrices is taken into account.  Of course, at
finite $N$ there are simple identities relating traces.
If one wants to take these
identities into account the derivation of the appropriate Hamiltonian 
is much more difficult\cite{jevics}\ even though the 
identities in \refe{ident}\ are still valid.   
{\it A priori} it is not clear that $\hat H$ is
suitable for finite $N.$  

While $\hat H$ certainly does reproduce formally the Schwinger-Dyson  equations,
it is also crucial for the existence of the stochastic quantization 
that $\hat H$ should have a unique ground state.  In 
perturbation theory in $1/N^{2}$ this is obvious.  Our aim is now to 
consider what happens at finite values of $N.$ 
We will show that requiring a unique vacuum and no negative energy
states implies relationships among the {\it a priori} independent creation
operators, and that $N$ be an integer.
These relations exactly enforce 
the known identities between traces of powers of matrices.  

We first consider a limit where this analysis simplifies.
A very important feature of this \sftH\ is that the string 
splitting and joining terms are independent of the choice of the gauge 
theory action.  Only the $\la$ dependent terms, which generate
deformations of loops and annihilation of small loops, depend on $S.$ 
In a precise sense, the choice of $S$ is equivalent to the choice of a 
background in which the strings propagate. Let us define
$H\equiv \lim_{\la\uparrow\infty}\hat H.$ Then
\begin{equation}
\exp\left(-{N^{2}\over\la}(\aaa1+\aaa{-1})\right) H\exp\left( 
{N^{2}\over\la}(\aaa1+\aaa{-1})\right)=\hat H.
\label{similarity}
\end{equation}
Thus a background, in this model, amounts to a coherent state
of Wilson loops. On this background other Wilson loops
`propagate' using the Hamiltonian $H.$

Let us start with the case $\la =\infty.$
The total winding number of the Wilson loops in the 
correlation function is preserved by $H.$ Acting with $H$ on 
a state with only positive windings   will only
produce states with positive windings, and the same is true for states 
with purely negative windings.  
We shall begin with this case. Later
we shall consider states with positive and negative winding strings.
We compute first in  the  winding number 
$w=2$ sector. There are two states  $a_{2}^{\dagger}|0\raa, 
(a_{1}^{\dagger})^{2}|0\raa$ corresponding to one doubly wound Wilson 
line and a product of two singly wound Wilson lines.  In this sector
$$H_{2}=\left(\matrix{2&2/N^{2}\cr {2}&2\cr}\right)$$
which has a vanishing eigenvalue if $N=1.$

In the $w=3$ sector, the states are $\aaa3|0\raa,$ $\aaa2\aaa1|0\raa,$ 
and $\aap13|0\raa.$ Here we find
$$H_{3}=\left(\matrix{3&4/N^{2}&0\cr 6&3&6/N^{2}\cr 
0&2&3\cr}\right).$$
For $N=2$ this has a vanishing eigenvalue with eigenvector
$$V_{3} \equiv \left(\aaa3 -3\aaa2\aaa1 + 2 \aap13\right)|0 \raa.$$
For $N>2$ there are no zero eigenvalues.  

In the $w=4$ sector, the states are $\aaa4|0\raa,\aaa3\aaa1|0\raa,$ 
$\aaa2\aap12|0\raa,$ $\aap22|0\raa,$ and $\aap14|0\raa.$
Now 
$$H_{4} =\left(\matrix{4&6/N^{2}&8/N^{2}&0&0\cr 8&4&0&8/N^{2}&0\cr
4&0&4&2/N^{2}&0\cr 0&6&4&4&12/N^{2}\cr 0&0&0&2&4\cr}\right).$$
This has a negative  eigenvalue for $N=2,$ a zero 
eigenvalue for $N=3$ and only positive eigenvalues for $N>3.$
The eigenvalues in a sector with winding number $w$ are of 
the form $w \pm (m/N)$ where $m$ is an integer.  For example, for
$w=2,m=2,$ for
$w=3,$ $m=0,6$ and for $w=4,$ $m=0,4,12.$  The eigenvalue of interest 
for a given value of $N$ 
is the largest value of $m,$ which is $w(w-1).$

The pattern evident in these examples 
continues: For a given value of $N,$ $H_{w}$ has a 
vanishing eigenvalue when $w=N+1,$ and negative eigenvalues if $w>N+1.$   
We see therefore that the stochastic quantization of $H$ is 
ill-defined as it stands. 
At the first level we demand that the vacuum is unique so we impose 
(for $N=2$) $V_{3}=0,$ {\it i.e.} we demand the identity
\begin{equation}
\aaa3 = 3\aaa2\aaa1 - 2 \aap13.
\end{equation}
This is just the  $2 \times 2$ matrix  identity
$$ \Tr U^{3} = 3\Tr U^{2}\Tr U - 2 (\Tr U)^{3}.$$
We see that this relation is 
encoded in the Hamiltonian automatically.   

If this is a constraint it has to be compatible with time
evolution. For $\la =\infty$ this is trivial in the $w=3$ sector. 
However, 
if we want now to go to the $w=4$ sector (keeping $N=2$) we must check
the consistency
of imposing this constraint on those states. 
Looking at states with $w=4$ at $N=2$ the consistency condition is
\begin{equation}
H\aaa3\aaa1|0 \raa= H(3\aaa2\aap12 - 2 \aap14)|0 \raa 
\end{equation}
but since we know
\begin{eqnarray}
H\aaa3\aaa1|0 \raa &=& \left(6 \aap12\aaa2 + 
{3\over 2} \aaa4\right)|0 \raa\nonumber\\
H\aap12\aaa2|0 \raa &=& \left(2\aap14 + {1\over 2} \aap22 + 
2\aaa3\aaa1\right)|0 \raa\nonumber\\
H\aap14|0 \raa &=& 3\aap12\aaa2|0 \raa ,
\end{eqnarray}
we must have a new constraint
\begin{equation}
V_{4}\equiv \left(\aaa4 - \aap22 - 4\aap12\aaa2 + 4\aap14\right)|0\raa = 0.
\end{equation}
This is precisely the identity
$$\Tr U^{4} - (\Tr U^{2})^{2} - 4(\Tr U)^{2} \Tr U^{2} + 4(\Tr 
U)^{4} = 0$$
valid for $2\times 2$ matrices.

\def\vac{|0\raa}
\def\co#1{{\cal O}_{#1}}  
To systematize what we have found, let us 
rephrase the manner in which $V_{4}=0$ was derived above.  Let 
$\co3\equiv\aaa3 -3\aaa2\aaa1 + 2 \aap13.$ We compute 
the commutator
$$\left[H, \aaa1\co3\right]_{+}= {3\over 2} \co4 - 2\aaa1\co3,$$
where $[]_{+}$ means we normal order the terms and keep only  terms with
creation operators, and $\co4\equiv \aaa4 - \aap22 - 4\aap12\aaa2 + 
4\aap14.$  
We found a
vanishing eigenvalue at $w=3$ when $N=2.$  For a unique vacuum, we must
impose the constraint $V_{3}\equiv 
\co3\vac=0,$ and since $HV_{3}=[H,\co3]\vac=0,$ we have 
$ \exp(-H\tau)V_{3}=0 $ for any $\tau.$  
However, this is not enough for 
consistency.  We must also ensure that 
$$ \exp(-H\tau) \co3 \prod_{i} \aaa{n_{i}}\vac=0$$
for all choices of $n_{i}$ and any $\tau.$  For example, this implies 
$ \exp(-H\tau)\co3 \aaa1\vac = 0$ for all $\tau.$  We must have
$$\left[H,\co3 \aaa1\right]_{+}\vac = \left[H,\co3 \aaa1\right]\vac = 
0,$$
which is precisely a linear combination of the new constraint
$\co4\vac=0$ and the initial constraint  $\co3 \aaa1\vac = 0$.  
We cannot stop here.  Clearly, $\exp(-\tau H)\co4\vac=0$ must be
true for all $\tau$ as well so we compute
\begin{equation}
[H,\co4]\vac = 2\co4\vac.
\end{equation}
Thus, as desired, $\co4\vac=0$ can be maintained without further
constraints.  This procedure proceeds recursively.  It is important
that at each winding number $w$ we get 
only one constraint. 

One can work out the constraints for positive and negative windings 
independently.  It appears to be possible to implement all these
constraints for any value of $N$ (not necessarily an integer)
as long as we do not consider
states with positive {\it and} negative windings.  However, 
explicit calculations show  that the 
constraints are only valid on all states 
when $N$ is an integer.  As an example
consider $ \aaa{-1}\co3\vac .$  For general $N$ 
$$\co3(N) = \aaa3-(3N/2)\aaa1\aaa2 + (N^{2}/2)\aap13$$
 satisfies
$H\co3(N)\vac = (3-6/N)\co3(N)\vac.$ We then find 
\begin{equation}
[H,\aaa{-1}\co3] \vac =  (4-6/N)\aaa{-1}\co3(N)\vac +{3\over 
N^{2}}(N-2)(\aaa2+N\aap12)\vac.
\end{equation}
Thus, only when $N=2$ is the constraint $\aaa{-1}\co3(N)\vac=0$  consistent 
with the time evolution defined by $H.$

We now take  $\la$ to be finite.  
$\hat H$ does not preserve winding number so  
time evolution takes us out of the winding sector we started with, so
we need  to check that the constraints are satisfied under this different time
evolution. This can be seen to follow from the relationship
in equation (\ref{similarity}). As a check we can compute
\begin{equation}
	[\hat H, \co{N+j}\prod_{i}\aaa{n_{i}}]\vac = 0.
\end{equation}
This is not surprising from the gauge theory perspective since 
the relations between traces are independent of the gauge theory 
action $S.$

Summarizing these calculations, we find that the Hamiltonian $\hat H$
when treated non-perturbatively at finite $N$ requires the 
integrality of $N.$  
For a given value of $N,$ we find an infinite set of 
$N$-dependent constraints on physical states   $\{\co{j}(N), 
j=N+1,\ldots\}$ which are 
consistently propagated by $\hat H.$  These constraints ensure that the 
vacuum is unique and that no negative energy states appear, and are
exactly the known relationships among traces of powers of 
$N \times N$ matrices.

We have taken some care in this analysis to avoid {\it any} use of 
the gauge theory origins of $\hat H.$  Indeed in the gauge theory the 
integrality of $N,$ the 
appearance of constraints, their form, and the fact that the 
constraints are background independent, are all completely obvious.  
{}From the string Hamiltonian point of view, we know of 
no obvious reason why the string field theory Hamiltonian  derived without 
regard to finite $N$ niceties exhibits all the required phenomena when 
treated carefully at finite values of the coupling, but in fact we 
have shown that it does! 

The Hamiltonian $H$ is  universal: it is 
part of {\it any} string theory derived from 
a \ymt.  Consider a given Wilson loop  specified as the holonomy of 
the gauge field around some curve $C.$  There are Wilson loops entirely 
analogous to the multiple winding states we considered above obtained 
by computing the holonomy for multiple circuits around $C.$  The 
joining and splitting of these loops is described precisely by $H.$
Thus the truncation of the string spectrum is universal in such 
Yang-Mills strings.  To be explicit, the point is that oscillator 
states of a multiply-wound string are identified with oscillator 
states of two (or more) strings at finite coupling.  The stronger the 
coupling, the fewer the independent states of a single string---this 
is a fascinating  dynamical phenomenon from the string
point of view.  It is difficult to see how such identifications of the 
string spectrum would be visible in first-quantized string theory, 
though we cannot exclude this definitively since we have only a  gauge-fixed 
framework\cite{v1}.

\section{Remarks and  speculations}
In what follows, we list some remarks and speculations.
\begin{itemize}
\item In this paper we started from a Hamiltonian designed to reproduce the 
Schwinger-Dyson equations, without explicitly taking matrix identities into
account.  This resulted in a simple Hamiltonian\cite{v2}, \refe{ham},
as compared to the complicated 
Hamiltonian which arises if one attempts to take the matrix identities into 
account explicitly\cite{jevics}.  What we found is that the simple Hamiltonian 
`dynamically' generates constraints that exactly implement the
matrix identities.  This of course results in a reduction of single string
states above winding $N\equiv 1/g_{\rm st}.$
\item The analogues of Wilson loops for theories with adjoint 
scalars are obvious  and satisfy the same constraints.  The naive
counting of dimensions suggests that the string theory constructed 
from $\cn=4,d=4$ \ymt\ is associated with a string theory in 11 
dimensions.  However, given a string field theory Hamiltonian in a 
particular gauge one has to explicitly calculate to deduce  
the approximate background geometry   on which
the string propagates, if such a classical geometry exists.  
In stochastic quantization only the infinite time 
correlation functions are uniquely defined\cite{pw}. This corresponds in 
the string theory to boundary correlation functions. 
We speculate that in the $\cn=4,d=4$  \ymt\ the stochastic
time will be identified with the radial coordinate of the $AdS_5,$ and that
this connection  is at the root of holography\cite{holo}.  It is by no means
clear that a holographic relationship exists in every case of
the suggested association\cite{v1}\ of a string theory with a Yang-Mills theory
even though the boundary correlation functions will  always be reproduced.
\item $q$-deformations have been suggested\cite{sanjaye}\  as 
explanations for the 
truncation of the spectrum of chiral primaries.  It would be 
interesting to see if they are related to our calculations.   In a 
different vein,  't Hooft has suggested that quantum gravity is a 
dissipative deterministic system\cite{thh}.  There may be a connection 
with the approach described in this paper.
\item Finally multiply wound strings appear  in perturbative calculations of
high energy string
scattering\cite{mende}.  It would be interesting to reconsider the
true high-energy behaviour in Yang-Mills strings
taking into account a truncation of the number of windings as we have found.

\end{itemize}
\section{Acknowledgements}
We thank H. Verlinde for  discussions.  This work was 
supported in part by NSF grant PHY98-02484.
\def\np#1#2#3{Nucl. Phys. B#1,  #3 (#2)}
\def\prd#1#2#3{Phys. Rev. D#1, #3 (#2)}
\def\prl#1#2#3{Phys. Rev. Lett. #1, #3 (#2)}
\def\pl#1#2#3{Phys. Lett. B#1, #3 (#2)}


\begin{thebibliography}{99}
	
\bibitem{sm} S. Mandelstam, Ann. of Phys. 19, 25(1962); Phys. Rev. 
175, 1580 (1968) 

\bibitem{others} J.--L. Gervais and A. Neveu, 
\pl{80}{1979}{255}; Y. Nambu, \pl{80}{1979}{372}; A. Polyakov, 
\pl{82}{1979}{247} 

\bibitem{loop} G. De Angelis, D. De Falco and F. Guerra, Nuovo Cim. 
Lett. 19, 55 (1977); F. Guerra, R. Marra and G. Immirzi, Nuovo Cim. 
Lett. 23, 237 (1978); E. Corrigan and B. Hasslacher, 
\pl{81}{1979}{181}; L. Durand and E. Mendel, \pl{85}{1979}{241}; 
D. Foerster, Phys. Lett. 87B, 83 (1979); T. Eguchi,
Phys. Lett. 87B, 91 (1979); Yu. Makeenko 
and A. Migdal, Phys. Lett. 88B, 135 (1979)
\bibitem{jevics} A. Jevicki and B. Sakita, Nucl. Phys. B185, 89 (1981)
 
\bibitem{pol} A. Polyakov, Nucl. Phys. Proc. Supp. 68, 1 (1998);
Int. J. Mod. Phys. A14, 645 (1999) 
\bibitem{thooft} G. 't Hooft, \np{72}{1974}{461}
\bibitem{juan} J. Maldacena, Adv. Theor. Math. Phys. 2 (1998) 231 
\bibitem{gkp}S. Gubser, I. Klebanov and A. Polyakov, \pl{428}{1998}{105}
\bibitem{wmalda} E. Witten, Adv. Theor. Math. Phys. 2 (1998) 253

 
\bibitem{v1} V. Periwal, {\sl String field theory Hamiltonians from 
Yang--Mills theories}, hep-th/9906052
\bibitem{marc} G. Marchesini, Nucl. Phys. B191, 214 (1981); B239,
135 (1984)

\bibitem{pw} G. Parisi and Y.-S. Wu, Sci. Sin. 24, 484 (1981)

\bibitem{jevic} A. Jevicki and J. Rodrigues, \np{421}{278}{1994}; see 
also S. Das and A. Jevicki, Mod. Phys. Lett. A5 (1990) 1639; 
G. Moore, N. Seiberg and M. Staudacher, \np{362}{1991} 665
 
 
\bibitem{ik} N. Ishibashi and H. Kawai, \pl{314}{1993}{190}; 
M. Fukuma, N. Ishibashi, H. Kawai and M. Ninomiya,
\np{427}{1994}{139}; M. Ikehara, N. Ishibashi, H. Kawai, T. Mogami, R. 
Nakayama and N. Sasakura, \prd{50}{1994}{7467};
N. Ishibashi and H. Kawai, \pl{322}{1994}{67}; 
\pl{352}{1995}{75}

\bibitem{v2} V. Periwal, {\sl A toy model of Polyakov duality}, 
hep-th/9908203
\bibitem{rod} J. Rodrigues, \np{260}{1985}{350}

\bibitem{holo} G. 't Hooft, in {\it Salamfestschrift: a collection of 
talks}, World Scientific Series in 20th century physics, v. 4, eds. A. Ali 
et al. (World Sci., 1993) and in Proc. Symp. {\it The Oskar Klein 
Centenary}, ed. U. Lindstr\"om (World Sci., 1995); L. Susskind, J. Math. 
Phys. 36, 6377 (1995)

\bibitem{sanjaye} A. Jevicki and S. Ramgoolam,  J. High Energy
Phys.  9904, 032 (1999); P.-M. Ho, S. Ramgoolam and R. Tatar, {\sl Quantum 
spacetimes and finite $N$ effects in 4d super Yang-Mills theories},
hep-th/9907145  

\bibitem{thh} G. 't Hooft, {\sl Quantum gravity as a dissipative 
deterministic system}, gr-qc/9903084


\bibitem{mende} D. Gross and P. Mende, \pl{197}{1987}{129}; 
\np{303}{1988}{407}

 

\end{thebibliography}
\end{document}